 \documentclass[final,5p,times,twocolumn,authoryear]{elsarticle}

\usepackage{amssymb}
\usepackage{lipsum}
\journal{Biomedical Signal Processing and Control}

\begin{document}

\begin{frontmatter}

%% Title, authors and addresses

%% use the tnoteref command within \title for footnotes;
%% use the tnotetext command for theassociated footnote;
%% use the fnref command within \author or \affiliation for footnotes;
%% use the fntext command for theassociated footnote;
%% use the corref command within \author for corresponding author footnotes;
%% use the cortext command for theassociated footnote;
%% use the ead command for the email address,
%% and the form \ead[url] for the home page:
%% \title{Title\tnoteref{label1}}
%% \tnotetext[label1]{}
%% \author{Name\corref{cor1}\fnref{label2}}
%% \ead{email address}
%% \ead[url]{home page}
%% \fntext[label2]{}
%% \cortext[cor1]{}
%% \affiliation{organization={},
%%            addressline={}, 
%%            city={},
%%            postcode={}, 
%%            state={},
%%            country={}}
%% \fntext[label3]{}

\title{Machine learning for triage of strokes with large vessel occlusion\\ using photoplethysmography biomarkers}

%% use optional labels to link authors explicitly to addresses:
%% \author[label1,label2]{}
%% \affiliation[label1]{organization={},
%%             addressline={},
%%             city={},
%%             postcode={},
%%             state={},
%%             country={}}
%%
%% \affiliation[label2]{organization={},
%%             addressline={},
%%             city={},
%%             postcode={},
%%             state={},
%%             country={}}

%\author[first]{Author name}
\author[1,2]{Márton Á. Goda}
\author[3]{Helen Badge}
\author[3]{Jasmeen Khan}
\author[2]{Yosef Solewicz}
\author[2]{Moran Davoodi}
\author[3]{Rumbidzai Teramayi}
\author[3]{Dennis Cordato}
\author[3]{Longting Lin} 
\author[3]{Lauren Christie}
\author[3]{Christopher Blair}
\author[3]{Gagan Sharma}
\author[3]{Mark Parsons}
\author[2]{Joachim A. Behar}

\affiliation[1]{organization={Pázmány Péter Catholic University Faculty of Information Technology and Bionics},
            addressline={Práter u. 50/A}, 
            city={Budapest},
            postcode={1083}, 
            state={},
            country={Hungary}}

\affiliation[2]{organization={Faculty of Biomedical Engineering, Technion Institute of Technology},
            addressline={Technion-IIT}, 
            city={Haifa},
            postcode={32000}, 
            state={},
            country={Israel}}

\affiliation[3]{organization={Ingham Institute for Applied Medical Research, Sydney Brain Center UNSW},
            addressline={Liverpool Hospital}, 
            city={Sydney},
            postcode={2052}, 
            state={},
            country={Australia}}

\begin{abstract}
\textit{Objective.} Large vessel occlusion (LVO) stroke presents a major challenge in clinical practice due to the potential for poor outcomes with delayed treatment. Treatment for LVO involves highly specialized care, in particular endovascular thrombectomy, and is available only at certain hospitals. Therefore, prehospital identification of LVO by emergency ambulance services, can be critical for triaging LVO stroke patients directly to a hospital with access to endovascular therapy. Clinical scores exist to help distinguish LVO from less severe strokes, but they are based on a series of examinations that can take minutes and may be impractical for patients with dementia or those who cannot follow commands due to their stroke. There is a need for a fast and reliable method to aid in the early identification of LVO. In this study, our objective was to assess the feasibility of using 30-second photoplethysmography (PPG) recording to assist in recognizing  LVO stroke. \textit{Method.} A total of 88 patients, including 25 with LVO, 27 with stroke mimic (SM), and 36 non-LVO stroke patients (NL), were recorded at the Liverpool Hospital emergency department in Sydney, Australia. Demographics (age, sex), as well as morphological features and beating rate variability measures, were extracted from the PPG. A binary classification approach was employed to differentiate between LVO stroke and NL+SM (NL.SM). A 2:1 train-test split was stratified and repeated randomly across 100 iterations. \textit{Results.} The best model achieved a median test set area under the receiver operating characteristic curve (AUROC) of 0.77 (0.71--0.82). \textit{Conclusion.} Our study demonstrates the potential of utilizing a 30-second PPG recording for identifying LVO stroke.
\end{abstract}

%%Graphical abstract
%\begin{graphicalabstract}
%\includegraphics{grabs}
%\end{graphicalabstract}

%%Research highlights
%\begin{highlights}
%\item Research highlight 1
%\item Research highlight 2
%\end{highlights}

\begin{keyword}
%% keywords here, in the form: keyword \sep keyword, up to a maximum of 6 keywords
large vessel occlusion \sep stroke, digital biomarkers \sep photoplethysmography \sep machine learning.

%% PACS codes here, in the form: \PACS code \sep code

%% MSC codes here, in the form: \MSC code \sep code
%% or \MSC[2008] code \sep code (2000 is the default)

\end{keyword}

\end{frontmatter}

%\tableofcontents

%% \linenumbers

%% main text

\section{Introduction}
\label{sec:introduction}
Stroke is the fifth leading cause of adult disability in the developed world \cite{Guzik_2017stroke}. Acute stroke is a medical emergency, especially in stroke caused by large vessel occlusions (LVO) \cite{Murray_2020artificial}. LVOs obstruct major cerebral arteries, and the decreased blood flow causes substantial brain damage and high rates of death and severe disability \cite{Krishnan_2021complications,Hendrix_2019risk}. Time-critical hyperacute LVO stroke treatments to restore cerebral blood flow, include mechanical (endovascular) thrombectomy and intravenous thrombolysis. Unfortunately thrombolysis is not effective at quickly clearing large clots, but endovascular thrombectomy is, and hence substantially reduces long-term, stroke disability \cite{Li_2023mechanical}. These benefits are time-dependent, with faster access to treatment associated with better outcomes \cite{Sheth_2015time}. However, there are often delays in the treatment that lead to poor outcomes \cite{Lachkhem_2018understanding}. 

Rapid early diagnosis of LVO stroke is the goal so that patients with LVO stroke are taken directly to thrombectomy-capable hospitals, and those with non-LVO stroke (NL) and stroke mimics (SM) that can be taken to the closest hospital, and not ‘overload’ the comprehensive stroke centers. However, prehospital assessment of LVO stroke remains challenging due to limited diagnostic accuracy with existing clinical measures used to predict the likelihood of LVO \cite{Smith_2018accuracy, Krebs_2018prehospital}. The National Institutes of Health Stroke Scale (NIHSS) \cite{Garcia_2023} and modified Rankin Scale (mRS)  serve distinct but complementary roles in the evaluation and management of LVO stroke \cite{Goyal_2016analysis}. The NIHSS is an 11-item clinical assessment designed to assess clinical deficits and stroke severity quickly, it can be completed pre-hospital and before imaging can be completed \cite{Smith_2018accuracy}. However, the time taken to administer the full scale is prohibitive in a prehospital environment, so an abbreviated version, the 8-item Hunter 8, has been validated as a valid and feasible measure for prehospital rapid assessment. Apart from limited specificity for diagnosing LVO vs NL or SM, another limitation of the various NIHSS Scores is that they cannot accurately be performed with patients with significant cognitive or communication impairments. The NIHSS has a sensitivity of 0.64–0.80 and a sensitivity of 0.72-0.84 for LVO stroke detection \cite{Smith_2018accuracy}. The area under the receiver operating characteristic curve (AUROC) for identifying LVO using the Hunter 8 was 0.73 (95\% CI 0.66-0.79) \cite{Garcia_2023hunter}. The limitations of these tools highlight the need for additional assessment methods for more accurate diagnosis of LVO stroke.

\begin{figure*}[!ht]
    \centering
    \includegraphics[width=\textwidth]{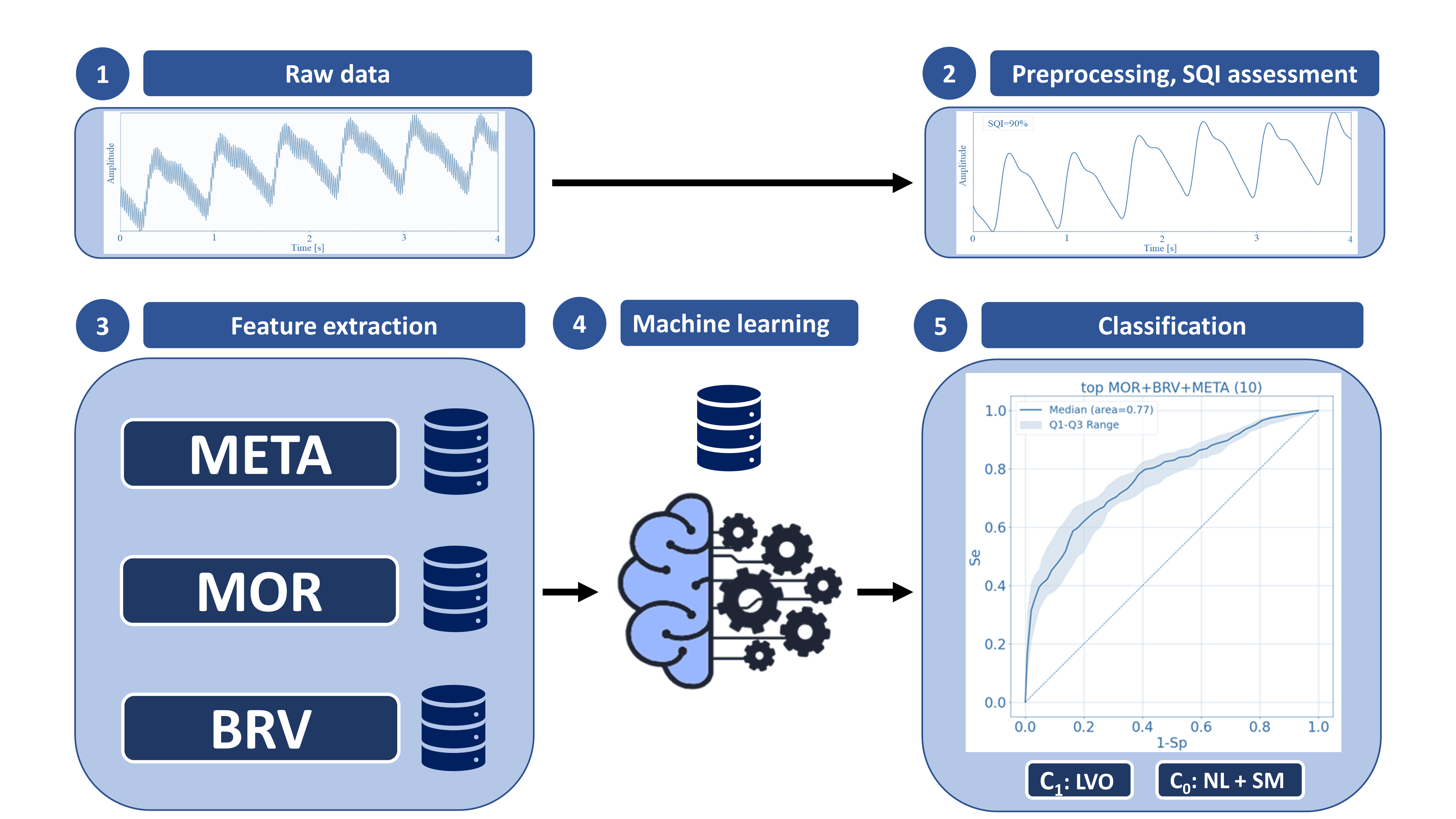}
    \caption{\scriptsize{Graphical Abstract. First, the process begins with raw PPG extraction. The second step is preprocessing, which includes filtering and signal quality assessment. In the third step, features are extracted, including   META features (age, sex), PPG morphological features (MOR), and beat rate variability (BRV). Fourth is the evaluation of feature importance, a machine learning-based model is trained and validated. Finally, in the fifth step, the triage of LVO stroke is tested and evaluated using area under the receiver-operating characteristics curve (AUROC).  LVO: Large vessel occlusion stroke. NL.SM: non-LVO stroke + stroke mimics.}}
    \label{fig:GraphAbs}
\end{figure*}

Strategies to enhance prehospital detection of LVO stroke include mobile stroke ambulances with qualified specialists and CT scanners onboard, that are effective at reducing delays to treatment. However, they are probably too expensive and resource-intensive for widespread implementation \cite{Navi_2022mobile}. Increasing the diagnostic capability for stroke in all ambulances would have greater reach and impact. Thus, in the future, ‘stroke-capable’ ambulances might have the capability to use multi-modal data routinely collected by paramedics, including clinical data, NIHSS and bio-signals such as electrocardiogram (ECG) and photoplethysmography (PPG) \cite{Yu_2022ai, Choi_2021deep}. Emerging evidence shows the potential for physiological data to predict LVO stroke. The PPG signal is an optical measurement of the arterial pulse wave \cite{Charlton2019, Charlton_2023wearable}, generated when blood is ejected from the heart  \cite{alastruey_arterial_2023}. The PPG signal can be obtained quickly using single-lead sensors with devices available on all ambulances. Based on the regulation of sympathetic and parasympathetic cardiovascular functions by the brain \cite{De_2019relationship}, we hypothesize that data-driven algorithms, trained using continuous raw PPG recordings, have the potential to detect structural changes indicative of stroke of LVO etiology. Furthermore, cardiac dysfunction can result from various neurologic injuries, including stroke and spinal cord injury \cite{Hu_2023brain}. Previous research has demonstrated the potential for using ML to predict LVO stroke based on physiological data. Accordingly, the main goal of this research is to evaluate the feasibility of triaging LVO stroke using a short-PPG recording (30 seconds long). Such a technology would enable rapid patient triage in ‘stroke smart ambulance’ emergency services.

\section{Materials and methods}\label{sec_method}
\subsection{Dataset}\label{method_database}

The study protocol received approval from the South West Sydney Human Ethics Committee (2019/ ETH00096 number). Data were collected at Liverpool Hospital in Sydney and measurements were taken in the emergency department. A \textit{Powerlab 16/35} (Ad Instruments, Bella Vista, NSW, Australia) device was used for the PPG measurements. The PPG sensors were placed on the left ear lobe and either on the left index finger or on the left middle finger. The finger pulse signal and the earlobe signal were measured at 5V. The sampling frequency for the PPG was set to 1 kHz. Overall, the fingertip PPG signal demonstrated superior quality; therefore, the fingertip PPG measurements were used for further analysis. 

\begin{figure}[!htb]
    \centering
    \includegraphics[width=\columnwidth]{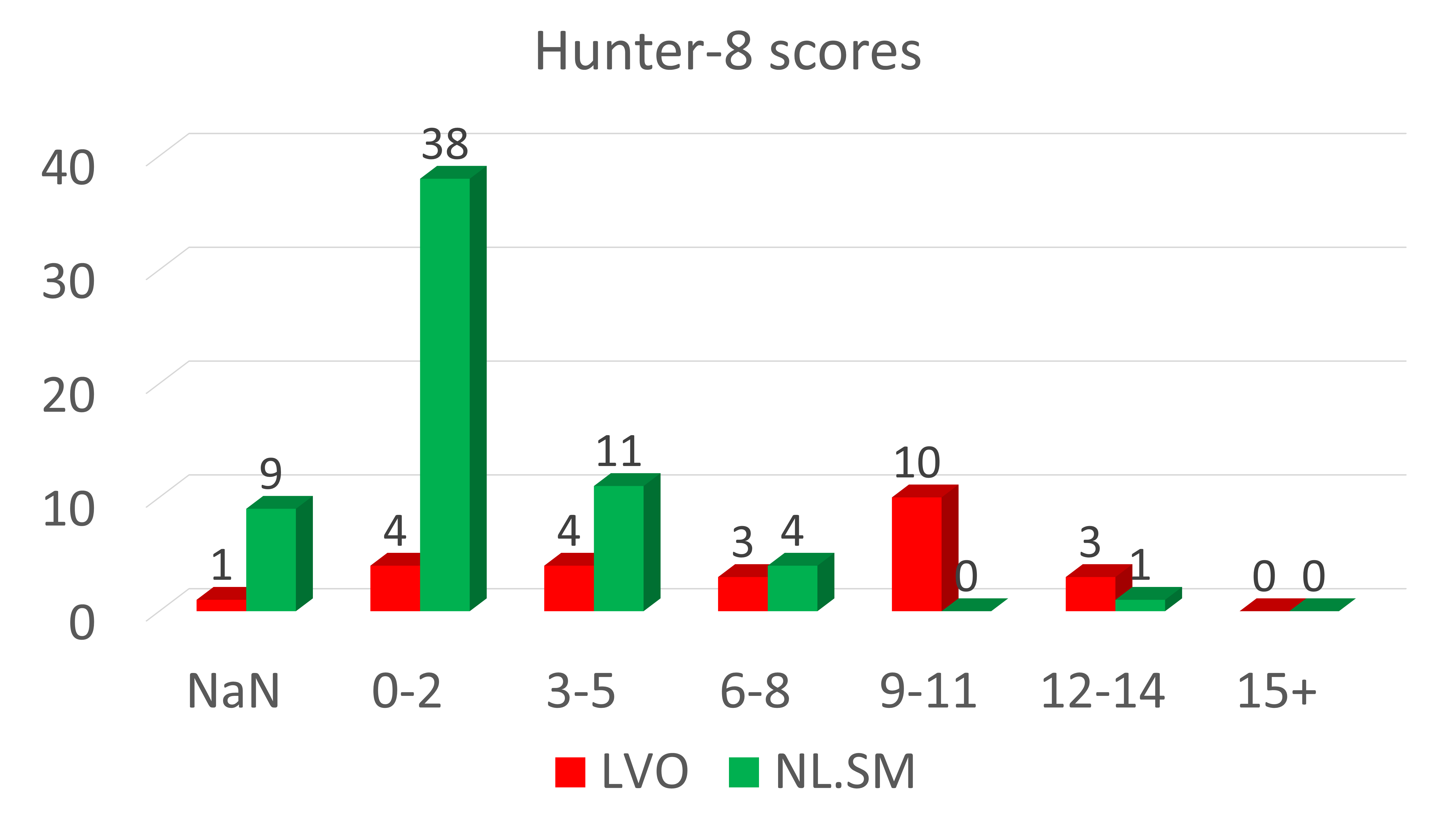}
    \caption{\scriptsize{Distribution of Hunter-8 scores. LVO: Large vessel
occlusion stroke. NL.SM: non-LVO stroke + stroke mimics.}}
    \label{fig:stroke_stat}
\end{figure}

\begin{figure}[!htb]
    \centering
    \includegraphics[width=0.85\columnwidth]{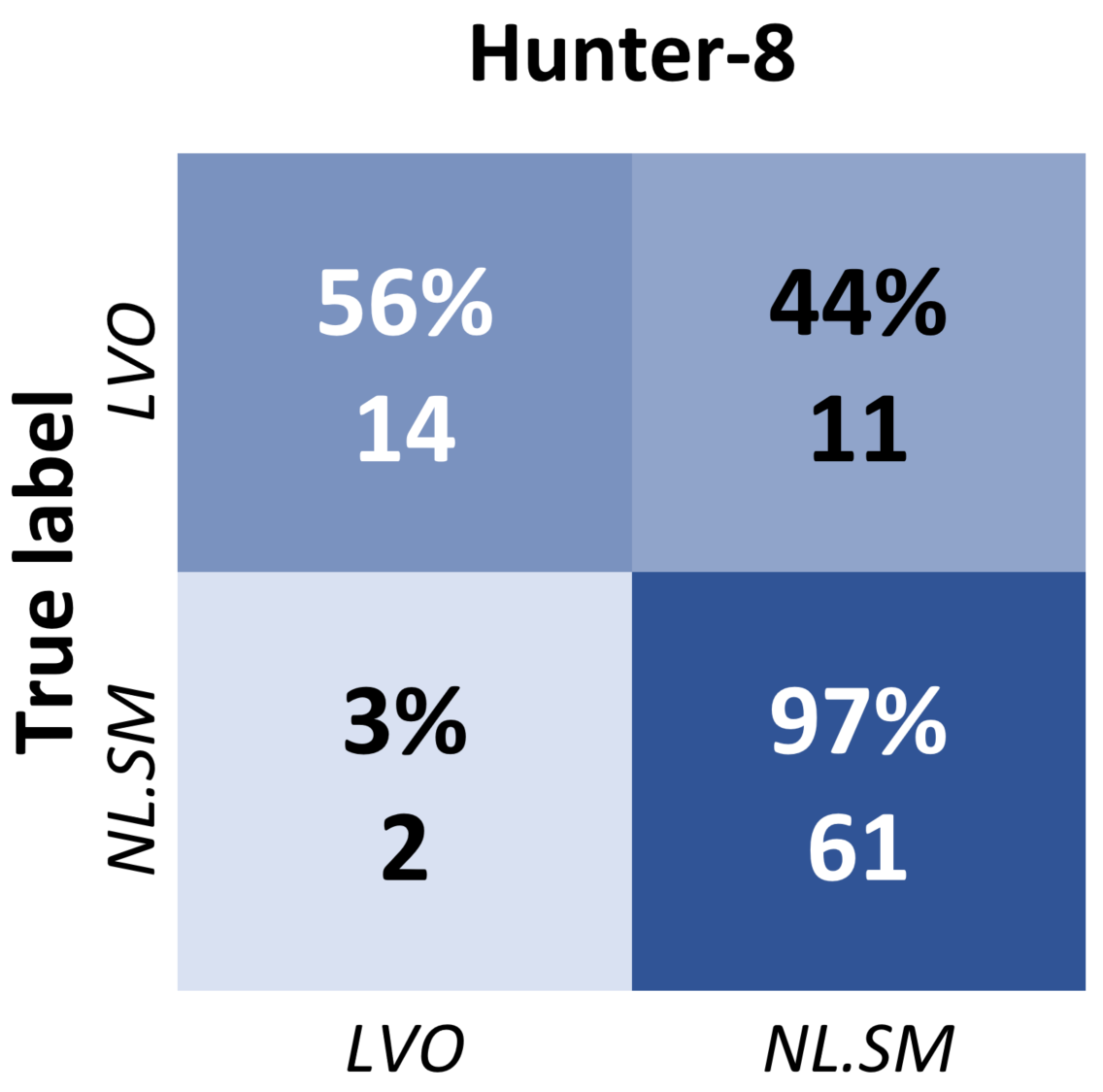}
    \caption{\scriptsize{Confusion Matrix for Hunter-8 Evaluation: This figure presents the classification performance of the Hunter-8 evaluation compared to confirmed diagnoses. The matrix highlights true positives, false positives, true negatives, and false negatives. Rows correspond to actual diagnoses, while columns represent predicted outcomes. LVO: Large vessel occlusion stroke. NL.SM: non-LVO stroke + stroke mimics.}}
    \label{fig:stroke_stat}
\end{figure}

Positive recordings for LVO (25 recordings) were classified as the positive class denoted as $C_1$, while the remaining 61 recordings, including LVO-mimic (SM) and non-strokes (NS), comprised the negative class denoted as $C_0$ (see Figure \ref{fig:stroke_stat} Panel A). SM presents an acute onset of focal neurological symptoms but is later diagnosed as having a non-vascular origin. The median duration of recordings was 10 minutes (interquartile range: 10-10 minutes), with a median age of 69.5 years (interquartile range: 61-77.3, see Figure \ref{fig:stroke_stat} Panel B), and 65\% of the participants were male. The median BMI was 27.5 (interquartile range: 24.4-31.1), with 13 patients lacking BMI data. Notably, approximately 60\% of the patients were either overweight or obese. The recordings were segmented into 30-second windows and each window was used as an example for training or testing.

\begin{figure}[!htb]
    \centering
    \includegraphics[width=\columnwidth]{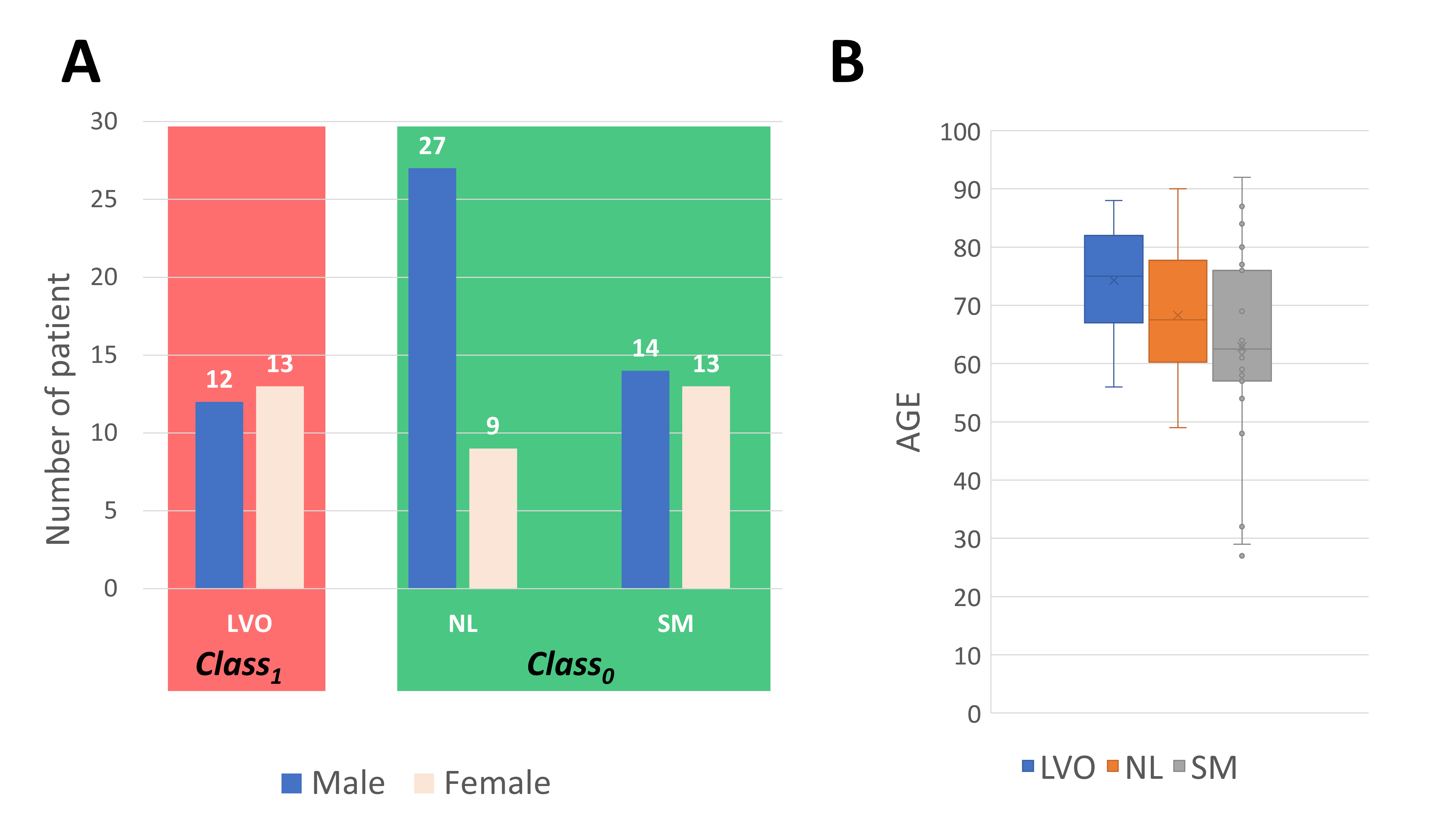}
    \caption{\scriptsize{Sex (Panel A) and age (Panel B) distribution by stroke types. LVO: Large vessel
occlusion stroke. NL: non-LVO stroke. SM: stroke mimics. NL.SM stroke includes both NL and SM cases.}}
    \label{fig:stroke_stat}
\end{figure}

\subsection{Preprocessing and feature extraction}
The raw PPG time series were pre-processed to eliminate baseline wander and high-frequency noise. This involved employing a zero-phase fourth-order Infinite Impulse Response (IIR) bandpass filter, set within the range of 0.5 to 12 Hz. Each recording was segmented into 30-second windows, totaling 1,801 windows for analysis. A total of 216 windows were excluded from the assessment because they were unsuitable for PPG signal quality index (SQI) calculation, primarily due to significant amplitude modulation. There was no information leakage to maintain data integrity, ensuring that all recordings for a given patient were assigned exclusively to either the training or test set. The PPG morphological (MOR) waveform features were extracted for each window using the $pyPPG$ Python toolboxe \cite{Goda_2024}, resulting in 101 MOR features. Additionally, Beating Rate Variability (BRV) features were derived based on the variability between consecutive PPG beats, resulting in 17 BRV features. Two META features, age, and sex were also used. During the feature calculation, the 10-minute recordings were divided into 30-second windows. For each MOR feature, the mean statistical descriptors were calculated within the 30-second window.

\begin{figure}[ht]
    \centering
    \includegraphics[width=0.95\columnwidth]{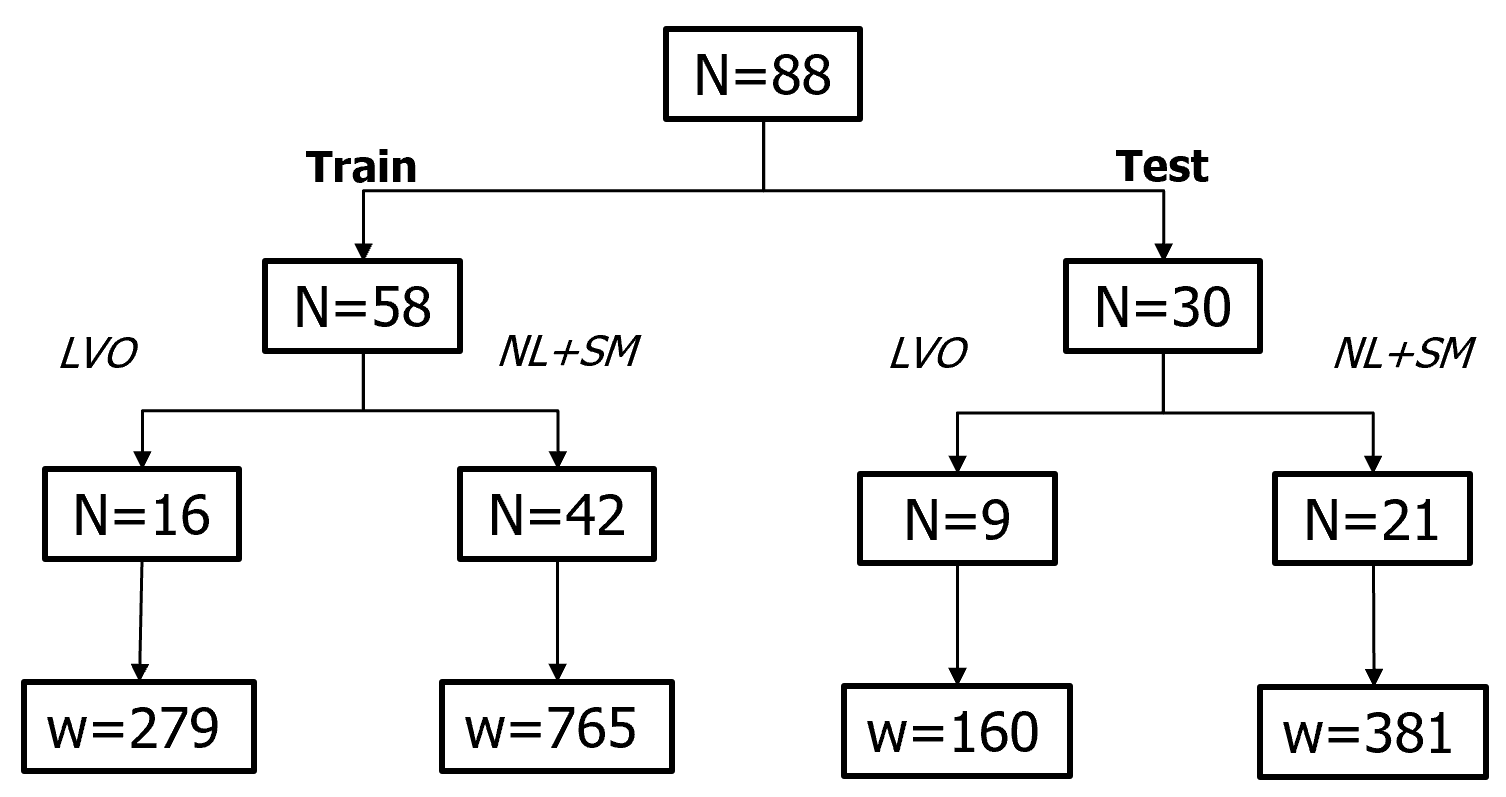}
    \caption{\scriptsize{Classification of 88 patients. Patients’ recordings (N) were divided into 30-second windows (w). The train and test sets are stratified following signal quality checks. LVO: Large vessel
occlusion stroke. NL: non-LVO stroke. SM: stroke mimics.}}
    \label{fig:trn-tst}
\end{figure}

\subsection{Machine learning}
The dataset was split into training and test sets, with the training set comprising 2/3 and the test set 1/3 of the entire dataset. This process was repeated 100 times with a random train-test split each time. On average, 9 patients were selected for the test set and 16 for the training set in $C_1$. For $C_0$, on average, 21 patients were chosen for the test set, while the remaining 42 were used for training. Three types of features were used: BRV, MOR, and META. Logistic regression models were trained when considering one of the feature types (i.e. BRV or MOR or META) and all feature types (BRV+MOR+META). This was done to evaluate the relative added value of each feature type. The top 10 features were selected based on their importance using recursive feature elimination.

\begin{table}[ht]
\centering
\caption{Top 10 feature names, descriptions, types, and units.}
\setlength{\tabcolsep}{3pt}
\renewcommand{\arraystretch}{1.5}
\small
\begin{tabular}{lp{5cm}ll}
\hline
\textbf{Feature} & \textbf{Description} & \textbf{Type} & \textbf{Unit} \\ \hline
$Age$ & age of the patient & META & year \\ \hline
$T_a$ & a-point time, the time between the pulse onset and a-point & MOR & s \\ \hline
$T_b$ & b-point time, the time between the pulse onset and b-point & MOR & s \\ \hline
$T_c$ & c-point time, the time between the pulse onset and c-point & MOR & s \\ \hline
$T_{dw25}/T_{sw25}$ & Ratio of the diastolic width vs. the systolic width at 25\% width & MOR & \% \\ \hline
$RMSSD$ & The RMSSD measure over a segment of peak-to-peak time series. & BRV & nu \\ \hline
$T_{b-d}$ & b-d time, the time between the b-point and d-point & MOR & s \\ \hline
$A_{p2}/A_{p1}$ & Ratio of the p2-point amplitude vs. the p1-point amplitude & MOR & \% \\ \hline
$AI$ & The ratio of the height of the late systolic peak to that of the early systolic peak in the pulse & MOR & \% \\ \hline
$T_{pw75}/T_{pi}$ & Ratio of the pulse width at 75\% of the systolic peak amplitude vs. the pulse interval & MOR & \% \\ \hline
\end{tabular}
\label{tab:features}
\end{table}

A total of 100 AUROC were computed from models using all four sets of features. Based on these AUROC values, the final results present the calculations of the median, 25$^{th}$ percentile, and 75$^{th}$ percentile.

\begin{figure}[ht]
    \centering
    \includegraphics[width=\columnwidth]{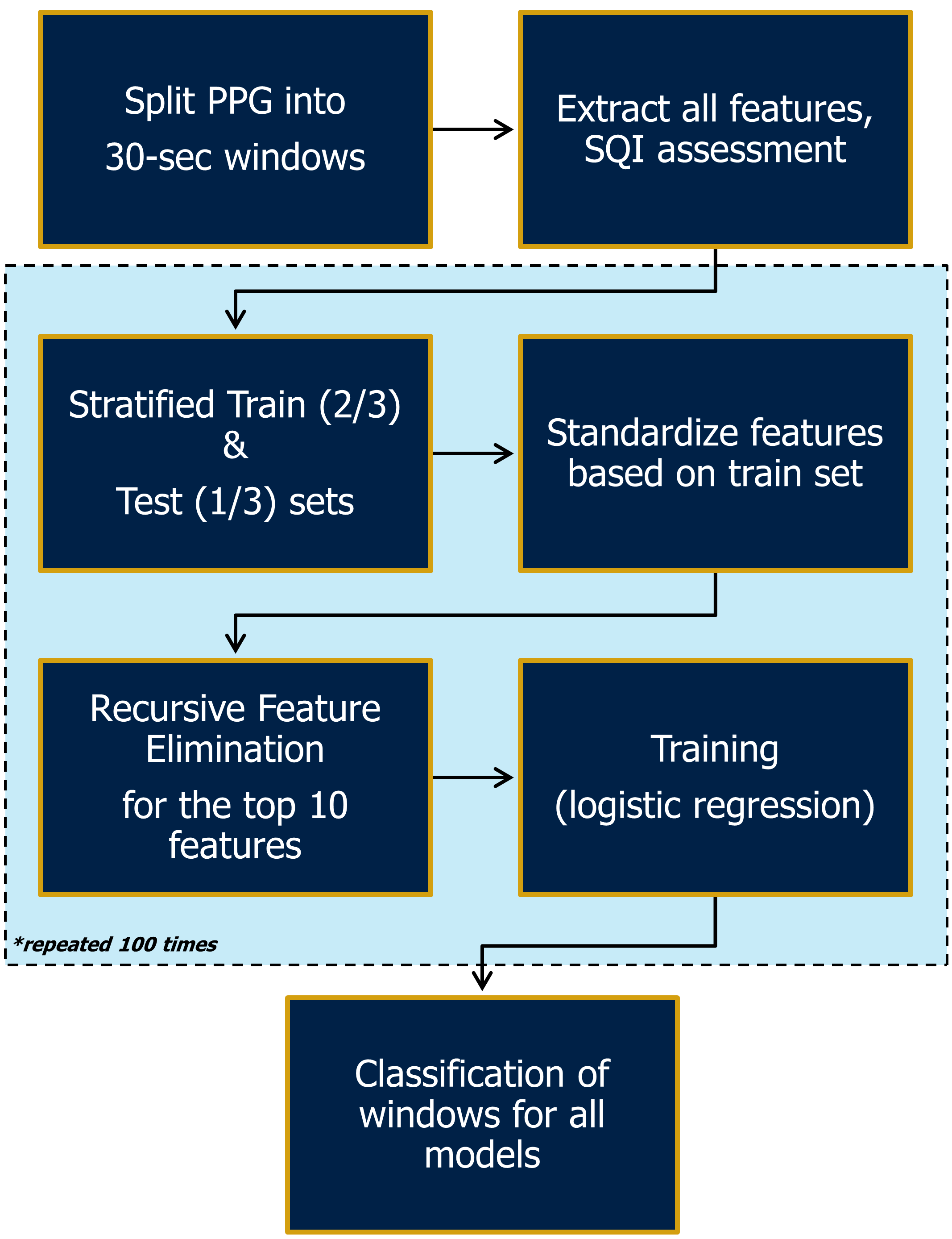}
    \caption{\scriptsize{Workflow of the training and evaluation process: The process begins with data preparation, involving the segmentation of 30-second PPG intervals and feature extraction. The training and validation phases are conducted iteratively 100 times on randomly shuffled yet stratified datasets to maintain class balance. Finally, model evaluation is performed to assess and compare performance metrics.}}
    \label{fig:pipeline}
\end{figure}

\subsection{Results}

\begin{figure}[ht]
    \centering
    \includegraphics[width=\columnwidth]{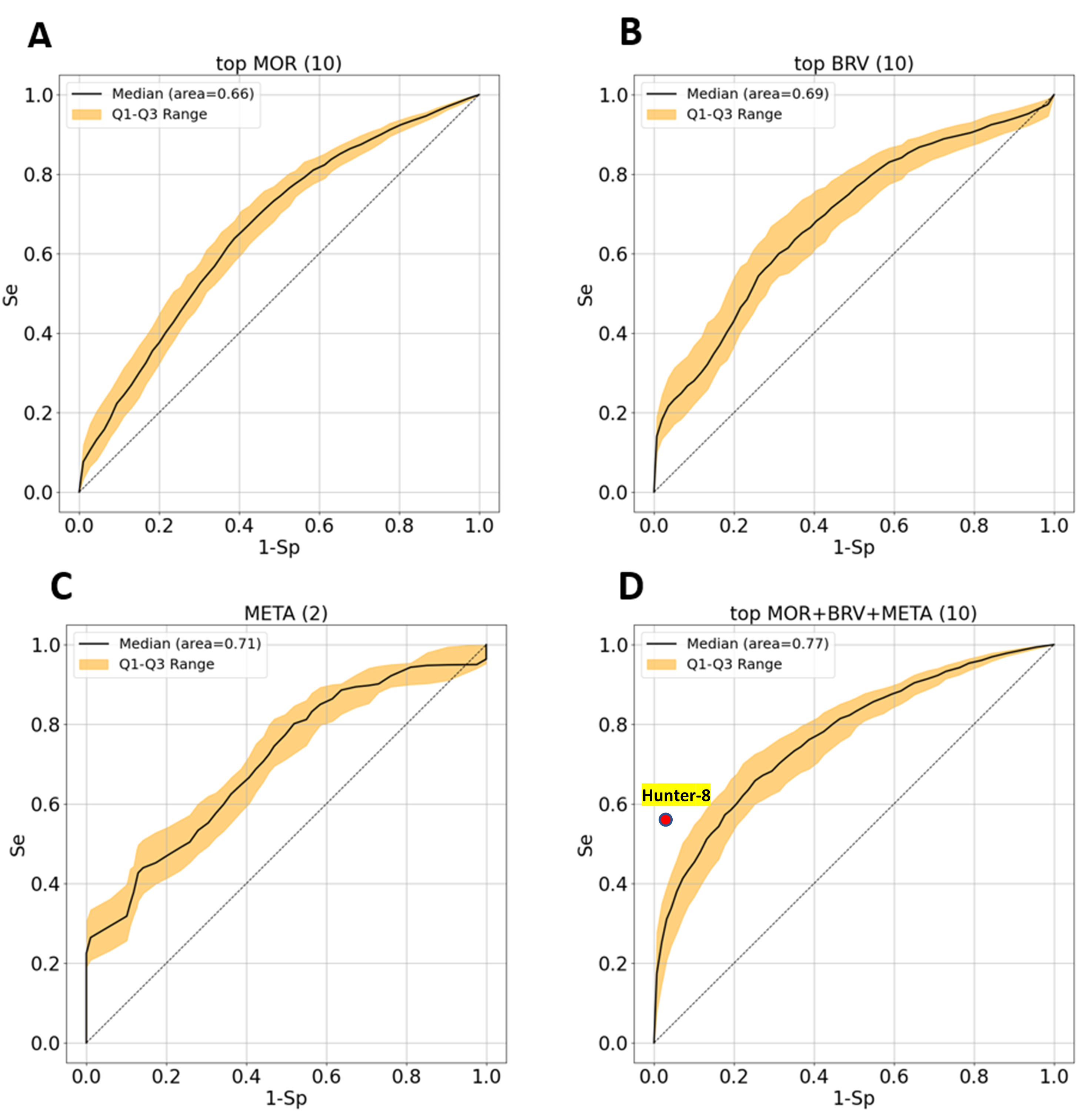}
    \caption{\scriptsize{Receiver operating characteristic curves for the per-patient classification task and as evaluated on the test sets, with the median (curve) and interquartile range (envelope). The models presented in Panels A, B, and C are trained with one type of feature, i.e. morphological feature (MOR), beat rate variability (BRV), or patient information (META), respectively. The model presented in Panel D is trained with MOR, BRV, and META features. The red point represents the sensitivity-specificity (SE-SP) cutoff for the Hunter-8 score used in clinical practice}. This point is evaluated for all patients who had a HUNTER-8 score documented in our cohort (76 out of 88 patients).}
    \label{fig:stroke_roc}
\end{figure}

Our project involved 88 patients, including 25 LVO and 61 NL.SM cases (see Figure \ref{fig:stroke_stat}). During training and validation, three types of features were used: META, MOR, and BRV. The best model utilized all feature types and achieved an AUROC of 0.77 (0.71--0.82) against 0.71 (0.64--0.77) for the best single modality model. For this model, the top 10 selected features were 1 META, 8 MOR, and 1 BRV. Table \ref{tab: results} summarizes the results of the LVO stroke triage. Figure \ref{fig:stroke_roc} presents the ROC curves for all the models. The following performance statistics were reported for the per-window classification task: the area under the ROC curve, precision, sensitivity, specificity, and F1 score.

\begin{figure}[ht]
    \centering
    \includegraphics[width=\columnwidth]{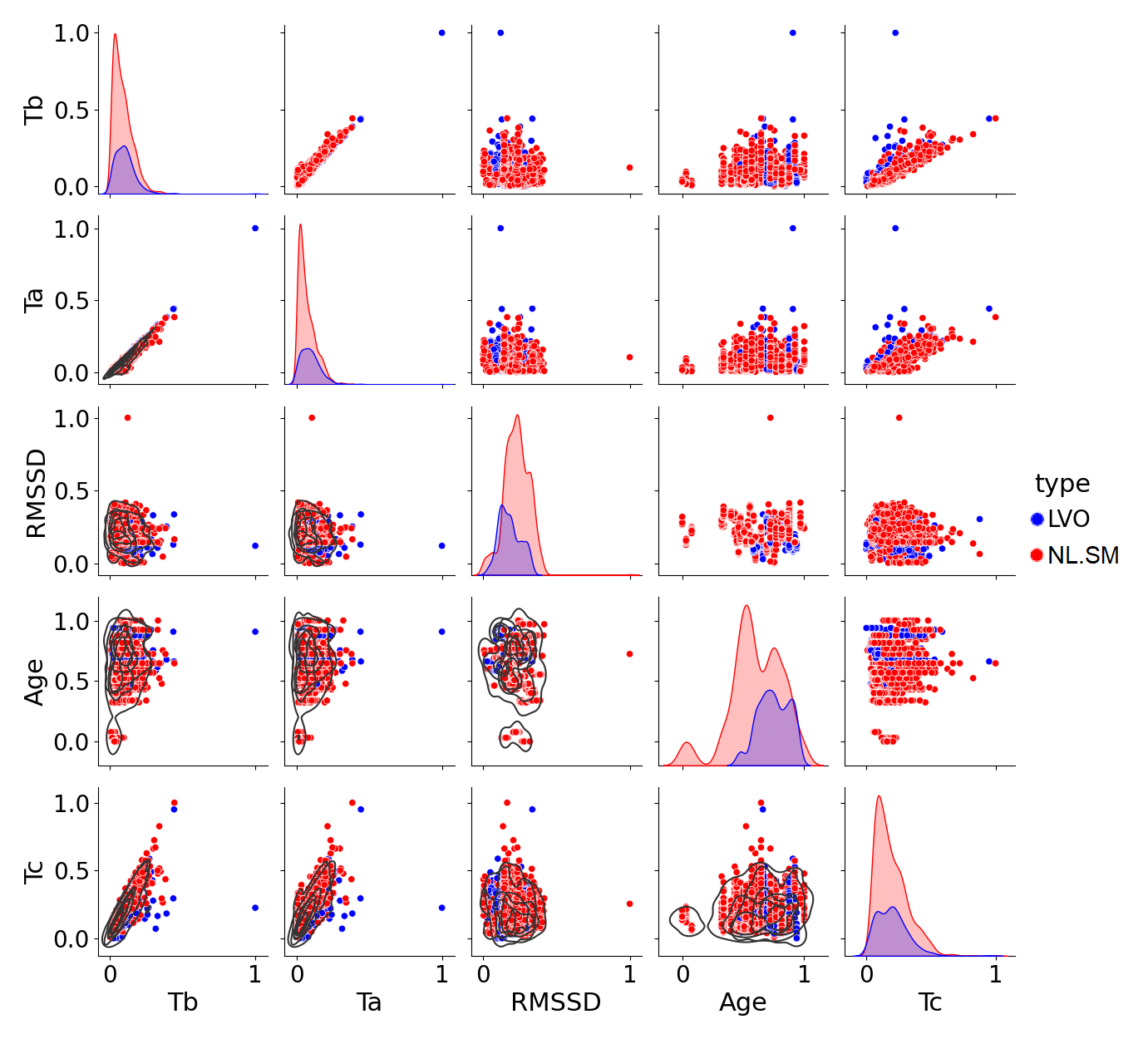}\caption{\scriptsize{Distribution plots for the top 5 most significant features, with blue indicating LVO type stroke and red indicating NL.SM, including non-LVO stroke (NL) and stroke mimics (SM) cases.}}
    \label{fig:feat_dist}
\end{figure}

In each iteration, the top 10 features were chosen for every trained model. To identify the most important features (across all trained models), we examined which features were most frequently selected among the top 10 across all 100 iterations. Specifically, the most frequently selected features are listed in the Table \ref{tab:features} \cite{Goda_2024, Charlot_2009interchangeability}. The top 5 features are presented in Figure \ref{fig:feat_dist}. This figure shows which features are most distinctive for the two classes. The diagonal plots represent the normalized distributions of each feature.

\begin{table}[ht!]
\caption{Results of LVO stroke triage.}
\centering
\setlength{\tabcolsep}{1pt}
\renewcommand{\arraystretch}{1.5}
\small
\begin{tabular}{cccccc}
\multicolumn{1}{c}{\textbf{Features}} & \multicolumn{1}{c}{\textbf{Sensitivity}} & \multicolumn{1}{c}{\textbf{Specificity}} & \multicolumn{1}{c}{\textbf{Precision}} & \multicolumn{1}{c}{\textbf{F1-score}} & \multicolumn{1}{c}{\textbf{AUROC}} \\ \hline
\textit{\textbf{MOR}} & 65\% & 59\%& 55\%& 68\% & 0.66 (0.62--0.71) \\
\textit{\textbf{BRV}} & 57\% & 72\% & 59\% & 68\% &0.69 (0.63--0.73)\\
\textit{\textbf{META}} & 41\% & 88\%& 59\%& 70\% &  0.71 (0.64--0.77)\\
\textit{\textbf{ALL}} & 66\%& 74\%&62\%  & 71\%& 0.77 (0.71--0.82) \\ \hline
\end{tabular}
\footnotetext{AUROC: area under the receiver operating characteristic curve. MOR: morphological features. \\BRV: Beat rate variability features. META: sex and, age. ALL: MOR+BRV+META features.}
\fontsize{7pt}{7pt}
\selectfont
\label{tab: results}
\end{table}

\section{Discussion}\label{sec_discuss}

This study demonstrated the potential of a novel data-driven approach using PPG signals to identify LVO stroke. The continuous physiological patient data recorded in the ambulance represents an opportunity for automated and fast stroke triage and could potentially support 'stroke-capable' ambulances in the future. We trained a machine learning model using PPG biomarkers to develop this clinical decision support tool. This data-driven approach presents an alternative to clinical scores that take longer and may be impractical for patients with dementia or those who cannot follow commands.

In 2023, Goda et al. \cite{Goda_2024} introduced a standardized $pyPPG$ toolbox for PPG feature extraction. The extracted PPG-based and META features were used for the first time in LVO stroke stratification. The main contribution of our study is demonstrating the feasibility of identifying strokes using short-term (30-second) PPG signals. Our analysis of feature importance demonstrated that the PPG-based features have a high predictive value. The current results introduce a novel medical assessment method utilizing short-term PPG signals. This innovative approach enhances the capability of smart ambulances, enabling rapid evaluation of LVO-type stroke orientation during patient hospitalization.

%In 2023, the classic manual LVO stroke ICH-at-risk expansion medical protocol \cite{Garcia_2023} achieved an AUROC of 0.73, which is lower than the performance of our PPG-based machine learning approach, achieved an AUROC of 0.77. 
The results obtained in this research are promising, indicating that data-driven based LVO stroke phenotyping holds significant potential for future applications using short-term (30-second) PPG recording. It is important to note that in our cohort, the HUNTER-8 demonstrates very high specificity ($\sim$0.96) but poor sensitivity ($\sim$0.57). The sensitivity is higher than previously published as the current dataset has more LVO stroke than would be expected in a random sample of pre-hospital patients as Liverpool Hospital is an endovascular center. Additionally, the Hunter-8 is unsuitable for patients who cannot respond to questions. This was the case for 13\% (12/88) of patients in this study, who had a missing HUNTER-8 score. Future research could explore using combined ECG data and whether multimodal clinical and physiological data will improve accurate diagnosis, along with AI-based voice, speech, and video recognition. Implementation studies to support integration into clinical practice will also be important.

The main limitation of this study is the small sample size. To enhance the accuracy and relevance of LVO stroke stratification, it is crucial to increase the dataset size and incorporate data recorded from ambulances. By including data captured during ambulance transport, the dataset will better reflect the patients' physiological status at the point of initial medical intervention. This approach can ensure a better understanding of patients' conditions as they are being taken by the ambulance, providing a clearer picture of their physiological state before they reach the emergency department. Further improvements may also be achieved using deep learning. Given the modest dataset size, we envisage the usage of a foundation model which would be fine-tuned to our downstream task. 

This study emphasized the feasibility of using a data-driven approach leveraging a short 30-second PPG recording to triage LVO strokes. It encourages more research and clinical exploration in this area to develop future technologies and smarter ambulances. By understanding the significance of PPG data, we can improve emergency medical services and help patients receive faster and more accurate stroke care.

\vspace{-0.5cm}
\section*{Acknowledgments}\label{sec_acknowledgments}
\vspace{-0.3cm}
MAG and JAB acknowledge the Estate of Zofia (Sophie) Fridman and funding from the Israel Innovation Authority. We acknowledge AusTechnion and the Ingham Institute for Applied Medical Research for their support in data collection. We acknowledge the assistance of ChatGPT, an AI-based language model developed by OpenAI, for its help in editing the English language of this manuscript.

\bibliographystyle{elsarticle-harv} 
\bibliography{refs}

\begin{thebibliography}{22}
\expandafter\ifx\csname natexlab\endcsname\relax\def\natexlab#1{#1}\fi
\providecommand{\url}[1]{\texttt{#1}}
\providecommand{\href}[2]{#2}
\providecommand{\path}[1]{#1}
\providecommand{\DOIprefix}{doi:}
\providecommand{\ArXivprefix}{arXiv:}
\providecommand{\URLprefix}{URL: }
\providecommand{\Pubmedprefix}{pmid:}
\providecommand{\doi}[1]{\href{http://dx.doi.org/#1}{\path{#1}}}
\providecommand{\Pubmed}[1]{\href{pmid:#1}{\path{#1}}}
\providecommand{\bibinfo}[2]{#2}
\ifx\xfnm\relax \def\xfnm[#1]{\unskip,\space#1}\fi
%Type = Article
\bibitem[{Alastruey et~al.(2023)Alastruey, Charlton, Bikia, Paliakait{\.e}, Hametner, Bruno, Mulder, Vennin, Piskin, Khir et~al.}]{alastruey_arterial_2023}
\bibinfo{author}{Alastruey, J.}, \bibinfo{author}{Charlton, P.H.}, \bibinfo{author}{Bikia, V.}, \bibinfo{author}{Paliakait{\.e}, B.}, \bibinfo{author}{Hametner, B.}, \bibinfo{author}{Bruno, R.M.}, \bibinfo{author}{Mulder, M.P.}, \bibinfo{author}{Vennin, S.}, \bibinfo{author}{Piskin, S.}, \bibinfo{author}{Khir, A.W.}, et~al., \bibinfo{year}{2023}.
\newblock \bibinfo{title}{Arterial pulse wave modelling and analysis for vascular age studies: a review from vascagenet}.
\newblock \bibinfo{journal}{American Journal of Physiology-Heart and Circulatory Physiology} .
%Type = Article
\bibitem[{Charlot et~al.(2009)Charlot, Cornolo, Brugniaux, Richalet and Pichon}]{Charlot_2009interchangeability}
\bibinfo{author}{Charlot, K.}, \bibinfo{author}{Cornolo, J.}, \bibinfo{author}{Brugniaux, J.V.}, \bibinfo{author}{Richalet, J.P.}, \bibinfo{author}{Pichon, A.}, \bibinfo{year}{2009}.
\newblock \bibinfo{title}{Interchangeability between heart rate and photoplethysmography variabilities during sympathetic stimulations}.
\newblock \bibinfo{journal}{Physiological measurement} \bibinfo{volume}{30}, \bibinfo{pages}{1357}.
%Type = Article
\bibitem[{Charlton et~al.(2023)Charlton, Allen, Bail{\'o}n, Baker, Behar, Chen, Clifford, Clifton, Davies, Ding et~al.}]{Charlton_2023wearable}
\bibinfo{author}{Charlton, P.H.}, \bibinfo{author}{Allen, J.}, \bibinfo{author}{Bail{\'o}n, R.}, \bibinfo{author}{Baker, S.}, \bibinfo{author}{Behar, J.A.}, \bibinfo{author}{Chen, F.}, \bibinfo{author}{Clifford, G.D.}, \bibinfo{author}{Clifton, D.A.}, \bibinfo{author}{Davies, H.J.}, \bibinfo{author}{Ding, C.}, et~al., \bibinfo{year}{2023}.
\newblock \bibinfo{title}{The 2023 wearable photoplethysmography roadmap}.
\newblock \bibinfo{journal}{Physiological measurement} \bibinfo{volume}{44}, \bibinfo{pages}{111001}.
%Type = Article
\bibitem[{Charlton et~al.(2019)Charlton, Mariscal~Harana, Vennin, Li, Chowienczyk and Alastruey}]{Charlton2019}
\bibinfo{author}{Charlton, P.H.}, \bibinfo{author}{Mariscal~Harana, J.}, \bibinfo{author}{Vennin, S.}, \bibinfo{author}{Li, Y.}, \bibinfo{author}{Chowienczyk, P.}, \bibinfo{author}{Alastruey, J.}, \bibinfo{year}{2019}.
\newblock \bibinfo{title}{Modeling arterial pulse waves in healthy aging: a database for in silico evaluation of hemodynamics and pulse wave indexes}.
\newblock \bibinfo{journal}{American Journal of Physiology-Heart and Circulatory Physiology} \bibinfo{volume}{317}, \bibinfo{pages}{H1062--H1085}.
%Type = Article
\bibitem[{Choi et~al.(2021)Choi, Park, Jun, Pyo, Cho, Lee and Yu}]{Choi_2021deep}
\bibinfo{author}{Choi, Y.A.}, \bibinfo{author}{Park, S.J.}, \bibinfo{author}{Jun, J.A.}, \bibinfo{author}{Pyo, C.S.}, \bibinfo{author}{Cho, K.H.}, \bibinfo{author}{Lee, H.S.}, \bibinfo{author}{Yu, J.H.}, \bibinfo{year}{2021}.
\newblock \bibinfo{title}{Deep learning-based stroke disease prediction system using real-time bio signals}.
\newblock \bibinfo{journal}{Sensors} \bibinfo{volume}{21}, \bibinfo{pages}{4269}.
%Type = Article
\bibitem[{de~la Cruz et~al.(2019)de~la Cruz, Schumann, K{\"o}hler, Reichenbach, Wagner and B{\"a}r}]{De_2019relationship}
\bibinfo{author}{de~la Cruz, F.}, \bibinfo{author}{Schumann, A.}, \bibinfo{author}{K{\"o}hler, S.}, \bibinfo{author}{Reichenbach, J.R.}, \bibinfo{author}{Wagner, G.}, \bibinfo{author}{B{\"a}r, K.J.}, \bibinfo{year}{2019}.
\newblock \bibinfo{title}{The relationship between heart rate and functional connectivity of brain regions involved in autonomic control}.
\newblock \bibinfo{journal}{Neuroimage} \bibinfo{volume}{196}, \bibinfo{pages}{318--328}.
%Type = Article
\bibitem[{Garcia-Esperon et~al.(2023a)Garcia-Esperon, Ostman, Walker, Chew, Edwards, Emery, Bendall, Alanati, Dunkerton, Starling~de Barros et~al.}]{Garcia_2023}
\bibinfo{author}{Garcia-Esperon, C.}, \bibinfo{author}{Ostman, C.}, \bibinfo{author}{Walker, F.}, \bibinfo{author}{Chew, B.}, \bibinfo{author}{Edwards, S.}, \bibinfo{author}{Emery, J.}, \bibinfo{author}{Bendall, J.}, \bibinfo{author}{Alanati, K.}, \bibinfo{author}{Dunkerton, S.}, \bibinfo{author}{Starling~de Barros, R.}, et~al., \bibinfo{year}{2023}a.
\newblock \bibinfo{title}{The hunter-8 scale prehospital triage workflow for identification of large vessel occlusion and brain haemorrhage}.
\newblock \bibinfo{journal}{Prehospital Emergency Care} \bibinfo{volume}{27}, \bibinfo{pages}{623--629}.
%Type = Article
\bibitem[{Garcia-Esperon et~al.(2023b)Garcia-Esperon, Ostman, Walker, Chew, Edwards, Emery, Bendall, Alanati, Dunkerton, Starling~de Barros et~al.}]{Garcia_2023hunter}
\bibinfo{author}{Garcia-Esperon, C.}, \bibinfo{author}{Ostman, C.}, \bibinfo{author}{Walker, F.}, \bibinfo{author}{Chew, B.}, \bibinfo{author}{Edwards, S.}, \bibinfo{author}{Emery, J.}, \bibinfo{author}{Bendall, J.}, \bibinfo{author}{Alanati, K.}, \bibinfo{author}{Dunkerton, S.}, \bibinfo{author}{Starling~de Barros, R.}, et~al., \bibinfo{year}{2023}b.
\newblock \bibinfo{title}{The hunter-8 scale prehospital triage workflow for identification of large vessel occlusion and brain haemorrhage}.
\newblock \bibinfo{journal}{Prehospital Emergency Care} \bibinfo{volume}{27}, \bibinfo{pages}{623--629}.
%Type = Article
\bibitem[{Goda et~al.(2024)Goda, Charlton and Behar}]{Goda_2024}
\bibinfo{author}{Goda, M.{\'A}.}, \bibinfo{author}{Charlton, P.H.}, \bibinfo{author}{Behar, J.A.}, \bibinfo{year}{2024}.
\newblock \bibinfo{title}{pyppg: A python toolbox for comprehensive photoplethysmography signal analysis}.
\newblock \bibinfo{journal}{Physiological Measurement} \bibinfo{volume}{45}, \bibinfo{pages}{045001}.
%Type = Article
\bibitem[{Goyal et~al.(2016)Goyal, Jadhav, Bonafe, Diener, Mendes~Pereira, Levy, Baxter, Jovin, Jahan, Menon et~al.}]{Goyal_2016analysis}
\bibinfo{author}{Goyal, M.}, \bibinfo{author}{Jadhav, A.P.}, \bibinfo{author}{Bonafe, A.}, \bibinfo{author}{Diener, H.}, \bibinfo{author}{Mendes~Pereira, V.}, \bibinfo{author}{Levy, E.}, \bibinfo{author}{Baxter, B.}, \bibinfo{author}{Jovin, T.}, \bibinfo{author}{Jahan, R.}, \bibinfo{author}{Menon, B.K.}, et~al., \bibinfo{year}{2016}.
\newblock \bibinfo{title}{Analysis of workflow and time to treatment and the effects on outcome in endovascular treatment of acute ischemic stroke: results from the swift prime randomized controlled trial}.
\newblock \bibinfo{journal}{Radiology} \bibinfo{volume}{279}, \bibinfo{pages}{888--897}.
%Type = Article
\bibitem[{Guzik and Bushnell(2017)}]{Guzik_2017stroke}
\bibinfo{author}{Guzik, A.}, \bibinfo{author}{Bushnell, C.}, \bibinfo{year}{2017}.
\newblock \bibinfo{title}{Stroke epidemiology and risk factor management}.
\newblock \bibinfo{journal}{CONTINUUM: Lifelong Learning in Neurology} \bibinfo{volume}{23}, \bibinfo{pages}{15--39}.
%Type = Article
\bibitem[{Hendrix et~al.(2019)Hendrix, Sofoluke, Adams, Kunaprayoon, Zand, Kolinovsky, Person, Gupta, Goren, Schirmer et~al.}]{Hendrix_2019risk}
\bibinfo{author}{Hendrix, P.}, \bibinfo{author}{Sofoluke, N.}, \bibinfo{author}{Adams, M.D.}, \bibinfo{author}{Kunaprayoon, S.}, \bibinfo{author}{Zand, R.}, \bibinfo{author}{Kolinovsky, A.N.}, \bibinfo{author}{Person, T.N.}, \bibinfo{author}{Gupta, M.}, \bibinfo{author}{Goren, O.}, \bibinfo{author}{Schirmer, C.M.}, et~al., \bibinfo{year}{2019}.
\newblock \bibinfo{title}{Risk factors for acute ischemic stroke caused by anterior large vessel occlusion}.
\newblock \bibinfo{journal}{Stroke} \bibinfo{volume}{50}, \bibinfo{pages}{1074--1080}.
%Type = Article
\bibitem[{Hu et~al.(2023)Hu, Abdullah, Nanna and Soufer}]{Hu_2023brain}
\bibinfo{author}{Hu, J.R.}, \bibinfo{author}{Abdullah, A.}, \bibinfo{author}{Nanna, M.G.}, \bibinfo{author}{Soufer, R.}, \bibinfo{year}{2023}.
\newblock \bibinfo{title}{The brain--heart axis: neuroinflammatory interactions in cardiovascular disease}.
\newblock \bibinfo{journal}{Current Cardiology Reports} \bibinfo{volume}{25}, \bibinfo{pages}{1745--1758}.
%Type = Article
\bibitem[{Krebs et~al.(2018)Krebs, Sharkey-Toppen, Cheek, Cortez, Larrimore, Keseg and Panchal}]{Krebs_2018prehospital}
\bibinfo{author}{Krebs, W.}, \bibinfo{author}{Sharkey-Toppen, T.P.}, \bibinfo{author}{Cheek, F.}, \bibinfo{author}{Cortez, E.}, \bibinfo{author}{Larrimore, A.}, \bibinfo{author}{Keseg, D.}, \bibinfo{author}{Panchal, A.R.}, \bibinfo{year}{2018}.
\newblock \bibinfo{title}{Prehospital stroke assessment for large vessel occlusions: a systematic review}.
\newblock \bibinfo{journal}{Prehospital Emergency Care} \bibinfo{volume}{22}, \bibinfo{pages}{180--188}.
%Type = Article
\bibitem[{Krishnan et~al.(2021)Krishnan, Mays and Elijovich}]{Krishnan_2021complications}
\bibinfo{author}{Krishnan, R.}, \bibinfo{author}{Mays, W.}, \bibinfo{author}{Elijovich, L.}, \bibinfo{year}{2021}.
\newblock \bibinfo{title}{Complications of mechanical thrombectomy in acute ischemic stroke}.
\newblock \bibinfo{journal}{Neurology} \bibinfo{volume}{97}, \bibinfo{pages}{S115--S125}.
%Type = Article
\bibitem[{Lachkhem et~al.(2018)Lachkhem, Rican and Minvielle}]{Lachkhem_2018understanding}
\bibinfo{author}{Lachkhem, Y.}, \bibinfo{author}{Rican, S.}, \bibinfo{author}{Minvielle, E.}, \bibinfo{year}{2018}.
\newblock \bibinfo{title}{Understanding delays in acute stroke care: a systematic review of reviews}.
\newblock \bibinfo{journal}{The European Journal of Public Health} \bibinfo{volume}{28}, \bibinfo{pages}{426--433}.
%Type = Article
\bibitem[{Li et~al.(2023)Li, Abdalkader, Siegler, Yaghi, Sarraj, Campbell, Yoo, Zaidat, Kaesmacher, Pujara et~al.}]{Li_2023mechanical}
\bibinfo{author}{Li, Q.}, \bibinfo{author}{Abdalkader, M.}, \bibinfo{author}{Siegler, J.E.}, \bibinfo{author}{Yaghi, S.}, \bibinfo{author}{Sarraj, A.}, \bibinfo{author}{Campbell, B.C.}, \bibinfo{author}{Yoo, A.J.}, \bibinfo{author}{Zaidat, O.O.}, \bibinfo{author}{Kaesmacher, J.}, \bibinfo{author}{Pujara, D.}, et~al., \bibinfo{year}{2023}.
\newblock \bibinfo{title}{Mechanical thrombectomy for large ischemic stroke: a systematic review and meta-analysis}.
\newblock \bibinfo{journal}{Neurology} \bibinfo{volume}{101}, \bibinfo{pages}{e922--e932}.
%Type = Article
\bibitem[{Murray et~al.(2020)Murray, Unberath, Hager and Hui}]{Murray_2020artificial}
\bibinfo{author}{Murray, N.M.}, \bibinfo{author}{Unberath, M.}, \bibinfo{author}{Hager, G.D.}, \bibinfo{author}{Hui, F.K.}, \bibinfo{year}{2020}.
\newblock \bibinfo{title}{Artificial intelligence to diagnose ischemic stroke and identify large vessel occlusions: a systematic review}.
\newblock \bibinfo{journal}{Journal of neurointerventional surgery} \bibinfo{volume}{12}, \bibinfo{pages}{156--164}.
%Type = Article
\bibitem[{Navi et~al.(2022)Navi, Audebert, Alexandrov, Cadilhac, Grotta and Group}]{Navi_2022mobile}
\bibinfo{author}{Navi, B.B.}, \bibinfo{author}{Audebert, H.J.}, \bibinfo{author}{Alexandrov, A.W.}, \bibinfo{author}{Cadilhac, D.A.}, \bibinfo{author}{Grotta, J.C.}, \bibinfo{author}{Group, P.P.S.T.O.W.}, \bibinfo{year}{2022}.
\newblock \bibinfo{title}{Mobile stroke units: evidence, gaps, and next steps}.
\newblock \bibinfo{journal}{Stroke} \bibinfo{volume}{53}, \bibinfo{pages}{2103--2113}.
%Type = Article
\bibitem[{Sheth et~al.(2015)Sheth, Jahan, Gralla, Pereira, Nogueira, Levy, Zaidat, Saver and Trialists}]{Sheth_2015time}
\bibinfo{author}{Sheth, S.A.}, \bibinfo{author}{Jahan, R.}, \bibinfo{author}{Gralla, J.}, \bibinfo{author}{Pereira, V.M.}, \bibinfo{author}{Nogueira, R.G.}, \bibinfo{author}{Levy, E.I.}, \bibinfo{author}{Zaidat, O.O.}, \bibinfo{author}{Saver, J.L.}, \bibinfo{author}{Trialists, S.S.}, \bibinfo{year}{2015}.
\newblock \bibinfo{title}{Time to endovascular reperfusion and degree of disability in acute stroke}.
\newblock \bibinfo{journal}{Annals of neurology} \bibinfo{volume}{78}, \bibinfo{pages}{584--593}.
%Type = Article
\bibitem[{Smith et~al.(2018)Smith, Kent, Bulsara, Leung, Lichtman, Reeves, Towfighi, Whiteley and Zahuranec}]{Smith_2018accuracy}
\bibinfo{author}{Smith, E.E.}, \bibinfo{author}{Kent, D.M.}, \bibinfo{author}{Bulsara, K.R.}, \bibinfo{author}{Leung, L.Y.}, \bibinfo{author}{Lichtman, J.H.}, \bibinfo{author}{Reeves, M.J.}, \bibinfo{author}{Towfighi, A.}, \bibinfo{author}{Whiteley, W.N.}, \bibinfo{author}{Zahuranec, D.B.}, \bibinfo{year}{2018}.
\newblock \bibinfo{title}{Accuracy of prediction instruments for diagnosing large vessel occlusion in individuals with suspected stroke: a systematic review for the 2018 guidelines for the early management of patients with acute ischemic stroke}.
\newblock \bibinfo{journal}{Stroke} \bibinfo{volume}{49}, \bibinfo{pages}{e111--e122}.
%Type = Article
\bibitem[{Yu et~al.(2022)Yu, Park, Kwon, Cho and Lee}]{Yu_2022ai}
\bibinfo{author}{Yu, J.}, \bibinfo{author}{Park, S.}, \bibinfo{author}{Kwon, S.H.}, \bibinfo{author}{Cho, K.H.}, \bibinfo{author}{Lee, H.}, \bibinfo{year}{2022}.
\newblock \bibinfo{title}{Ai-based stroke disease prediction system using ecg and ppg bio-signals}.
\newblock \bibinfo{journal}{IEEE Access} \bibinfo{volume}{10}, \bibinfo{pages}{43623--43638}.

\end{thebibliography}

%% else use the following coding to input the bibitems directly in the
%% TeX file.

%%\begin{thebibliography}{00}

%% \bibitem[Author(year)]{label}
%% For example:

%% \bibitem[Aladro et al.(2015)]{Aladro15} Aladro, R., Martín, S., Riquelme, D., et al. 2015, \aas, 579, A101

%%\end{thebibliography}

\end{document}